\documentclass[aps,prb,twocolumn,superscriptaddress,floatfix]{revtex4-1}
\usepackage{hyperref}
\usepackage{graphicx}
\pdfoutput=1
\usepackage{dcolumn}
\usepackage{bm}
\usepackage{amssymb}
\usepackage{microtype}
\usepackage{xfrac}

\newcommand{\pyro}[2]{#1$_2$#2$_2$O$_7$}

\begin{document}

\title{Long-wavelength correlations in ferromagnetic titanate pyrochlores as revealed by small angle neutron scattering}

\author{C.~R.~C.~Buhariwalla}
\affiliation{Department of Physics and Astronomy, McMaster University, Hamilton, ON, L8S 4M1, Canada}

\author{Q.~Ma}
\affiliation{Department of Physics and Astronomy, McMaster University, Hamilton, ON, L8S 4M1, Canada}

\author{L.~DeBeer-Schmitt}
\affiliation{Chemical and Engineering Materials Division, Oak Ridge National Laboratory, Oak Ridge, TN, 37831, USA}

\author{K.~G.~S.~Xie}
\affiliation{Department of Physics and Astronomy and Guelph-Waterloo Physics Institute, University of Waterloo, Waterloo, Ontario N2L 3G1, Canada}
\affiliation{Institute for Quantum Computing, University of Waterloo, Waterloo, Ontario N2L 3G1, Canada}

\author{D.~Pomaranski}
\affiliation{Department of Physics and Astronomy and Guelph-Waterloo Physics Institute, University of Waterloo, Waterloo, Ontario N2L 3G1, Canada}
\affiliation{Institute for Quantum Computing, University of Waterloo, Waterloo, Ontario N2L 3G1, Canada}

\author{J.~Gaudet}
\affiliation{Department of Physics and Astronomy, McMaster University, Hamilton, ON, L8S 4M1, Canada}

\author{T.~J.~Munsie}
\affiliation{Department of Physics and Astronomy, McMaster University, Hamilton, ON, L8S 4M1, Canada}

\author{H.~A.~Dabkowska}
\affiliation{Brockhouse Institute for Materials Research, Hamilton, ON L8S 4M1 Canada}

\author{J.~B.~Kycia}
\affiliation{Department of Physics and Astronomy and Guelph-Waterloo Physics Institute, University of Waterloo, Waterloo, Ontario N2L 3G1, Canada}
\affiliation{Institute for Quantum Computing, University of Waterloo, Waterloo, Ontario N2L 3G1, Canada}

\author{B.~D.~Gaulin}
\affiliation{Department of Physics and Astronomy, McMaster University, Hamilton, ON, L8S 4M1, Canada}
\affiliation{Canadian Institute for Advanced Research, 180 Dundas St. W., Toronto, ON, M5G 1Z7, Canada}
\affiliation{Brockhouse Institute for Materials Research, Hamilton, ON L8S 4M1 Canada}

\date{\today}

\begin{abstract}
We have carried out small angle neutron scattering measurements on single crystals of two members of the family of cubic rare-earth titanate pyrochlores that display ferromagnetic Curie-Weiss susceptibilities, \pyro{Yb}{Ti} and \pyro{Ho}{Ti}.  \pyro{Ho}{Ti} is established as displaying a prototypical classical dipolar spin ice ground state, while \pyro{Yb}{Ti} has been purported as a candidate for a quantum spin ice ground state.  While both materials have been well studied with neutron scattering techniques, neither has been previously explored in single crystal form with small angle neutron scattering (SANS).  Our results for \pyro{Yb}{Ti} show distinct SANS features below its $\Theta_{CW}$$\sim$ 0.50 K, with rods of diffuse scattering extending along $\langle 111 \rangle$ directions in reciprocal space, off-rod scattering which peaks in temperature near $\Theta_{CW}$, and quasi-Bragg scattering at very small angles which correlates well with T$_C$ $\sim$ 0.26 K.  The quasi-Bragg scattering corresponds to finite extent ferromagnetic domains $\sim$ 140 \AA~across, at the lowest temperatures.  We interpret the $\langle 111\rangle$ rods of diffuse scattering as arising from domain boundaries between the finite-extent ferromagnetic domains.  In contrast the SANS signal in \pyro{Ho}{Ti} is isotropic within the (HHL) plane around {\bf Q}=0.  However the strength of this overall SANS signal has a temperature dependence resembling that of the magnetic heat capacity, with a peak near 3 K.  Below the break between the field-cooled and the zero-field cooled susceptibility in \pyro{Ho}{Ti} at $\sim$ 0.60 K, the SANS signal is very low, approaching zero. 
 
\end{abstract}

\maketitle
\section{Introduction}

The rare-earth titanate family of cubic pyrochlores of the form \pyro{RE}{Ti} has been of great interest over more than 15 years.   The Ti$^{4+}$ site is non-magnetic, and these materials corresponds to RE$^{3+}$ ions, most of which are magnetic, decorating a network of corner-sharing tetrahedra.  This is the canonical crystalline architecture supporting geometrical frustration in three dimensions \cite{diep2013frustrated,greedan2006frustrated,gardner2010magnetic}.  The \pyro{RE}{Ti} family has been particularly well studied, as it possesses relatively low melting temperatures and can be easily grown as single crystals using floating zone image furnace techniques \cite{dabkowska2010crystal}.  The magnetic ground states displayed across this family are quite varied.  The members of the family tend to display different combinations of magnetic interactions and anisotropy, and it is this combination that conspires to determine the nature of the magnetic ground state they display.

There are only three members of the \pyro{RE}{Ti} family whose net interactions at the near-neighbor level are ferromagnetic: \pyro{Yb}{Ti}, \pyro{Ho}{Ti}, and \pyro{Dy}{Ti}.  Of these, \pyro{Ho}{Ti} and \pyro{Dy}{Ti} have large Ising magnetic moments and leading interactions between moments are driven by dipolar interactions, rather than exchange interactions.  Both materials have low temperature states known to be excellent realizations of the classical dipolar spin ice model\cite{harris97,ramirez1999zero,Bramwell2001,Bramwell2001a,den2000dipolar,isakov2005spin,rauging2015}.  The magnetic structure factor associated with the dipolar spin ice model differs from that of the spin ice model primarily around {\bf Q}=0.  This is because the ultimate consequence of the dipolar sum in this three dimensional system is to constrain any net magnetic moment, and any scattering at {\bf Q}=0 at low enough temperature.  \pyro{Ho}{Ti} is a much easier subject of a neutron scattering experiment compared with \pyro{Dy}{Ti}, due to constraints from strong neutron absorption associated with naturally occurring Dy. Nonetheless, measurements employing isotopic $^{162}$Dy have been successfully carried out~\cite{fennell2004neutron}. 

By contrast \pyro{Yb}{Ti} possesses both small moments, $\sim$ 2 $\mu_B$ and easy plane or XY anisotropy~\cite{blote1969heat,gaudet2015,Hodges2001a}.  The former tends to negate dipolar interactions while the latter gives rise to a crystal field ground state doublet comprised primarily of J$_z$=$\pm$ 1/2 eigenstates~\cite{gaudet2015}, and hence quantum spin degrees of freedom.  The microscopic spin Hamiltonian for \pyro{Yb}{Ti} has been estimated from inelastic neutron scattering~\cite{Ross2011,robert2015spin,thompson2017quasiparticle}, and it is known to be described by anisotropic exchange at the near-neighbor level. These studies have led to \pyro{Yb}{Ti} being put forward as a candidate for a quantum spin ice ground state~\cite{hermele2004,benton2012,shannon2012,applegate2012Vindication,hayre2013,gingras2014quantum,Savary2017} - a theoretical U(1) quantum spin liquid with an emergent quantum electrodynamics, and magnetic and electric monopoles as well as gauge photons as elementary excitations~\cite{hermele2004,benton2012}. The ground state has also been argued to be subject to multiphase competition arising from the nearby antiferromagnetic phases observed in other rare-earth pyrochlores~\cite{jaubert2015multiphase,han2017,robert2015spin,hallas2017experimental}. Nonetheless, the overall interactions in \pyro{Yb}{Ti} are ferromagnetic as it shows a robust Curie-Weiss susceptibility with $\Theta_{CW}$$\sim$ 0.5 K~\cite{bramwell2000bulk,blote1969heat,Hodges2001a}.

Ferromagnetic materials are very well suited to study with small angle neutron scattering (SANS) techniques, as this diffraction technique is optimized for resolution and intensity around {\bf Q}=0, a magnetic zone centre for any magnetic structure with a net magnetization.  For that reason, it may be surprising that such a SANS study has yet to be carried out on single crystals of the ferromagnetic \pyro{RE}{Ti} family members.  The {\bf Q}-resolution is very good and flexible, as it can be adjusted with both the incident wavelength and the sample-detector distance, and this also controls the dynamic range in {\bf Q} accessible in a measurement.   As a diffraction technique (there is no energy discrimination on the scattered neutrons), it gives an undistorted S({\bf Q}) when the range of the inelastic features being integrated over is small compared with the incident neutron energy.  \pyro{Yb}{Ti} is known to display diffuse scattering with a characteristic extent in energy of $\sim$ 0.3 meV~\cite{Ross2009,ross2011dimensional}, while the diffuse magnetic scattering in the classical spin ice state of \pyro{Ho}{Ti} is known to be very elastic, with an extent in energy well below 0.1 meV~\cite{clancy2009revisiting}.  Therefore both \pyro{Yb}{Ti} and \pyro{Ho}{Ti} are excellent candidates for SANS single crystal studies, the subject of this paper.

\section{Methods}

This program of SANS measurements on both \pyro{Ho}{Ti} and \pyro{Yb}{Ti} was carried on the GP-SANS instrument~\cite{GPSANS} in the HFIR guide hall of Oak Ridge National Laboratory.  A $\sim$ 7 gram single crystal of \pyro{Ho}{Ti} and a $\sim$ 3.5 gram single crystal of \pyro{Yb}{Ti} were grown using optical floating zone methods following standard protocols~\cite{gardner1998}. The \pyro{Ho}{Ti} was post-annealed in air for 24 hours to reduce oxygen defects, causing a color change from brown to a more transparent tan color. 

In order to characterize the \pyro{Yb}{Ti} sample, heat capacity measurements were performed on the entire single crystal used in SANS measurement.  The heat capacity measurements were performed using the quasi-adiabatic method~\cite{Quilliam2007} with a 1 k$\Omega$ RuO$_2$ thermometer and 10 k$\Omega$ heater mounted directly on the thermally isolated sample.  Fine Cu wire was used to provide a dominant weak thermal link to the cold plate of an adiabatic demagnetization refrigerator.  The time constant of relaxation was chosen to be approximately half an hour, in order to make the internal equilibration time of the sample much shorter and thus insignificant.  Due to the large mass of the sample, the addenda contributions to the specific heat are negligible.

The single crystals of \pyro {Ho}{Ti} and \pyro {Yb}{Ti} were aligned and attached to copper and aluminum posts respectively, and mounted in a dilution refrigerator.  Samples mounted on aluminium rods were cooled to low temperatures in the presence of a weak magnetic field produced by external permanent magnets, to improve equilibration of the samples below $\sim$ 1 K, where the aluminum rods are superconducting.  We employed an alignment with the incident neutron beam parallel to the [HH0] direction.  Data were collected using a sample to detector distance of 4~m, and an incident neutron wavelength of 4~\AA with~a $\frac{\delta \lambda}{\lambda}$ =0.15.  The position sensitive neutron detector employed had a 1 m$^2$ area, giving an effective $| Q |$ range from $| Q |=$ 0.02~\AA$^{-1}$ to 0.15~\AA$^{-1}$ for all measurements presented here. The sample was rocked $\pm 2^\circ$ around the vertical [00L] crystalline axis, with 5 minutes exposure for each 0.5$^\circ$ step, except in the case of \pyro{Yb}{Ti} in the presence of a magnetic field where the sample was similarly rocked around a vertical [{H$\bar{\rm{H}}$0}] direction parallel to the applied magnetic field. SANS data were analyzed using the GRASP software package~\cite{grasp}. 

\section{Results and Discussion}

\subsection{\pyro{Yb}{Ti}}

\begin{figure}[htb]
\linespread{1}
\par
\includegraphics[width=\linewidth]{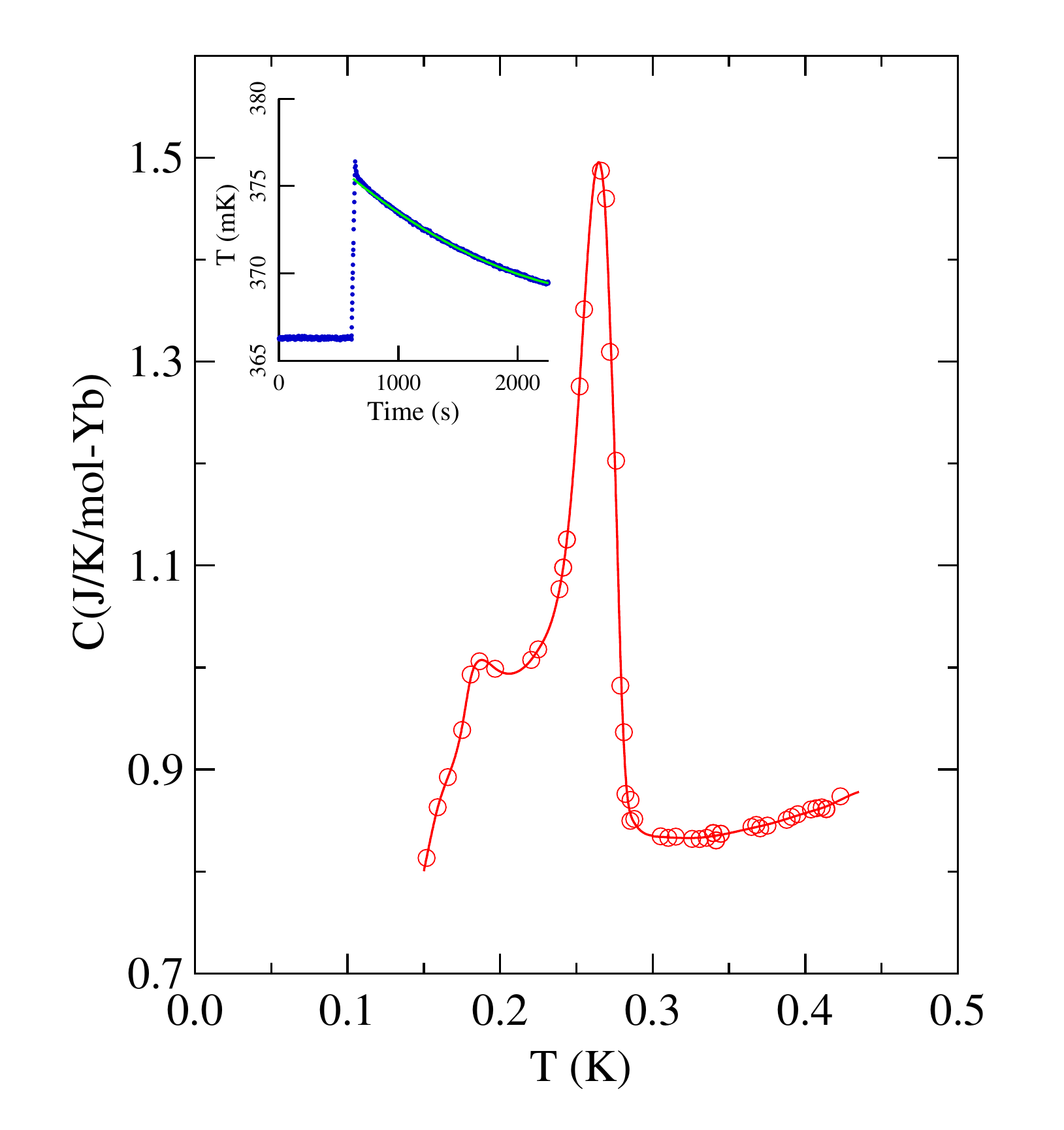}
\par
\caption{\emph{Low-temperature specific heat of single crystal \pyro{Yb}{Ti}.~ }  The heat capacity of the single crystal shows a sharp peak at 0.266 K with a second, broad and smaller feature at 0.185 K. The line is a guide to the eye. The inset shows a typical heat pulse used within the quasi-adiabatic method. }
\label{fig:yb_heatcap}
\end{figure}

\begin{figure*}[htb]
\linespread{1}
\par
\includegraphics[width=\linewidth]{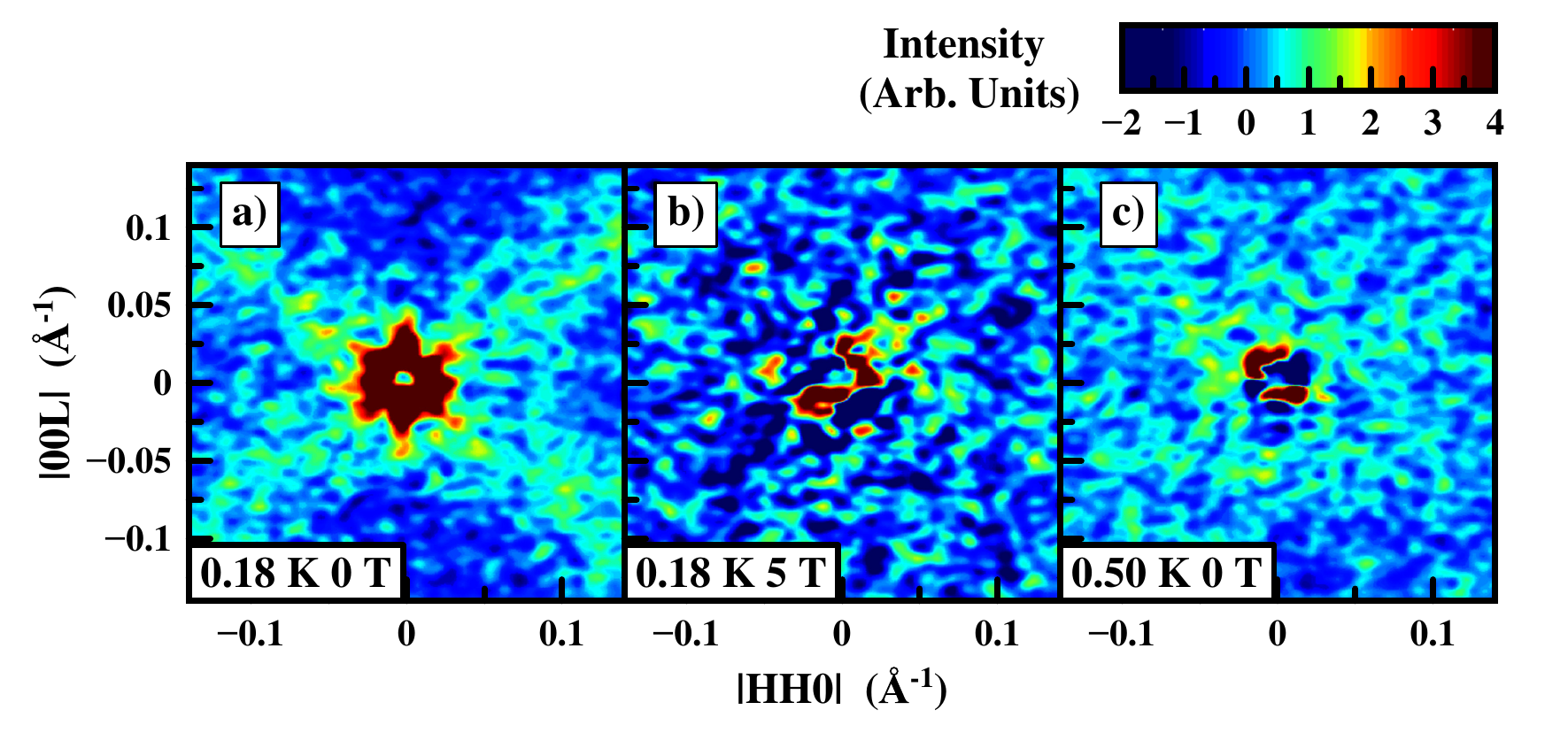}
\par
\caption{\emph{A comparison of low temperature, SANS data with corresponding data with a high, H=5 T, magnetic field applied along the {H$\bar{\rm{H}}$0} direction of \pyro{Yb}{Ti}.~ }  a)T=0.18~K, H = 0~T. b) T=0.180~K, H = 5~T. c) T=0.50~K, H = 0~T. Panels a) and c) have had a high temperature, T=0.80 K, data set subtracted while panel b) had a T=0.70 K high temperature data set subtracted. }
\label{fig:yb_panel}
\end{figure*}

The heat capacity of the large single crystal of \pyro{Yb}{Ti} that was the subject of the SANS experiment was measured using quasi adiabatic-techniques, and is shown in Fig.~\ref{fig:yb_heatcap}.  A typical heat pulse generating a temperature increase of~2\% is shown in the inset of Fig.~\ref{fig:yb_heatcap}.  In and of itself, a heat capacity measurement on such a large single crystal is unusual - most C$_P$ measurements are reported on very small crystals, which are less likely to display inhomogeneities.  In order to make contact between the SANS measurements and the heat capacity results we report, it is important to characterize the C$_P$ of the full single crystal employed in the SANS measurement.  

The details of the ground state in \pyro{Yb}{Ti} are known to be very sensitive to the precise stoichiometry of the material studied~\cite{ross2012lightly,DOrtenzio2013,arpino2017impact}, with single crystals tending to display weak ``stuffing", such that the actual stoichiometry of single crystals is Yb$_{2+x}$Ti$_{2-x}$O$_{7+\delta}$ with $x \sim$ 0.04.  This effect has been shown to cause the sharp T $\sim$0.265 K heat capacity anomaly correlated with a first order phase transition, to broaden and move to lower temperatures~\cite{arpino2017impact} with increasing $x$.  The heat capacity of our single crystal is characterized by a large, sharp C$_P$ anomaly at T$_C$= 0.266 K and a smaller broad anomaly at T= 0.185 K.  The sharp peak at 0.266 K indicates the presence of a dominant stoichiometric phase within the single crystal, while the smaller, broader peak at 185 mK is consistent with an approximate level of stuffing $x$ = + 0.01 and a smaller volume fraction.  

\pyro{Yb}{Ti} is known to display diffuse rods of scattering which extend along the $\langle 111\rangle$ directions in reciprocal space~\cite{bonville2004transitions}.  These rods of scattering have been extensively measured in single crystal samples using a variety of neutron techniques~\cite{Ross2009,thompson2011,chang2012higgs}, but only at large {\bf Q}, and their physical origin has been far from clear.  Earlier measurements tended to concentrate around the 111 Bragg position in reciprocal space, as the neutron instruments used were optimized for scattering at large {\bf Q}.  A subset of single crystals also show ferromagnetic Bragg scattering at the 111 position itself~\cite{yasui2003ferromagnetic,chang2012higgs}.  This Bragg position is an allowed reflection of the pyrochlore chemical structure, hence the magnetic contribution to the peak is small, $\sim$ 1 $\%$ of the total Bragg scattering even for a 1 $\mu_B$ ferromagnetic ordered moment.  The temperature dependence of Bragg scattering displays hysteresis, characteristic of a first order phase transition, with the $\Delta$T$_C$ observed ranging from 0.06 K~\cite{chang2012higgs} to 0.1 K~\cite{Scheie2017} for studies of the 111 and 2$\bar{2}$0 peaks respectively .

Should these diffuse rods of scattering extend down to {\bf Q}=0, they would be visible in a SANS experiment, and could be explored with substantially better {\bf Q} resolution than the previous measurements allowed.  We expected that this would be the case for several reasons:  the Curie-Weiss constant for \pyro{Yb}{Ti} is + 0.50 K, the mean field ordered phase predicted on the basis of the determination of the microscopic anisotropic exchange Hamiltonian gives a splayed ferromagnetic structure~\cite{Ross2011,han2017}, and a pronounced upturn in the temperature dependence of the uniform susceptibility of \pyro{Yb}{Ti} below $\sim$ 0.30 K has been measured~\cite{yasui2003ferromagnetic,lhotel2014}.

SANS measurements on our single crystal of \pyro{Yb}{Ti} are shown in Fig.~\ref{fig:yb_panel}.  Panels a) and c) show data taken in zero magnetic field at a) T=0.18 K, below both T$_C$ and $\Theta_{CW}$ and at c) T=0.50 K $\sim$ $\Theta_{CW}$.  A zero magnetic field data set at T=0.80 K has been used as a background for both a) and c).  Panel b) shows the corresponding data set at T=0.18 K, but with a H=5 T magnetic field applied along the 1$\bar{1}$0 direction of the crystal.  A background data set at T=0.70 K and H=5 T was employed for the data in Fig.~\ref{fig:yb_panel} c).  Four features are clear from inspection of Fig.~\ref{fig:yb_panel}.  First, rods of diffuse scattering extending out from {\bf Q}=0, along $\langle 111  \rangle$, are clearly visible in zero field at T=0.18 K.  Second, quasi-Bragg scattering is observed close to {\bf Q}=0 in zero magnetic field at T=0.18 K.  Third, both the diffuse scattering along $\langle 111\rangle$ directions and the quasi-Bragg scattering near {\bf Q}=0 are largely gone by T=0.50 K $\sim$ $\Theta_{CW}$.  Fourth, no substantial SANS signal is observed at low temperatures in the presence of an H=5 T magnetic field, where the ground state of the material is expected to be a long range ordered ferromagnet.

\begin{figure}[htb]
\linespread{1}
\par
\includegraphics[width=\linewidth]{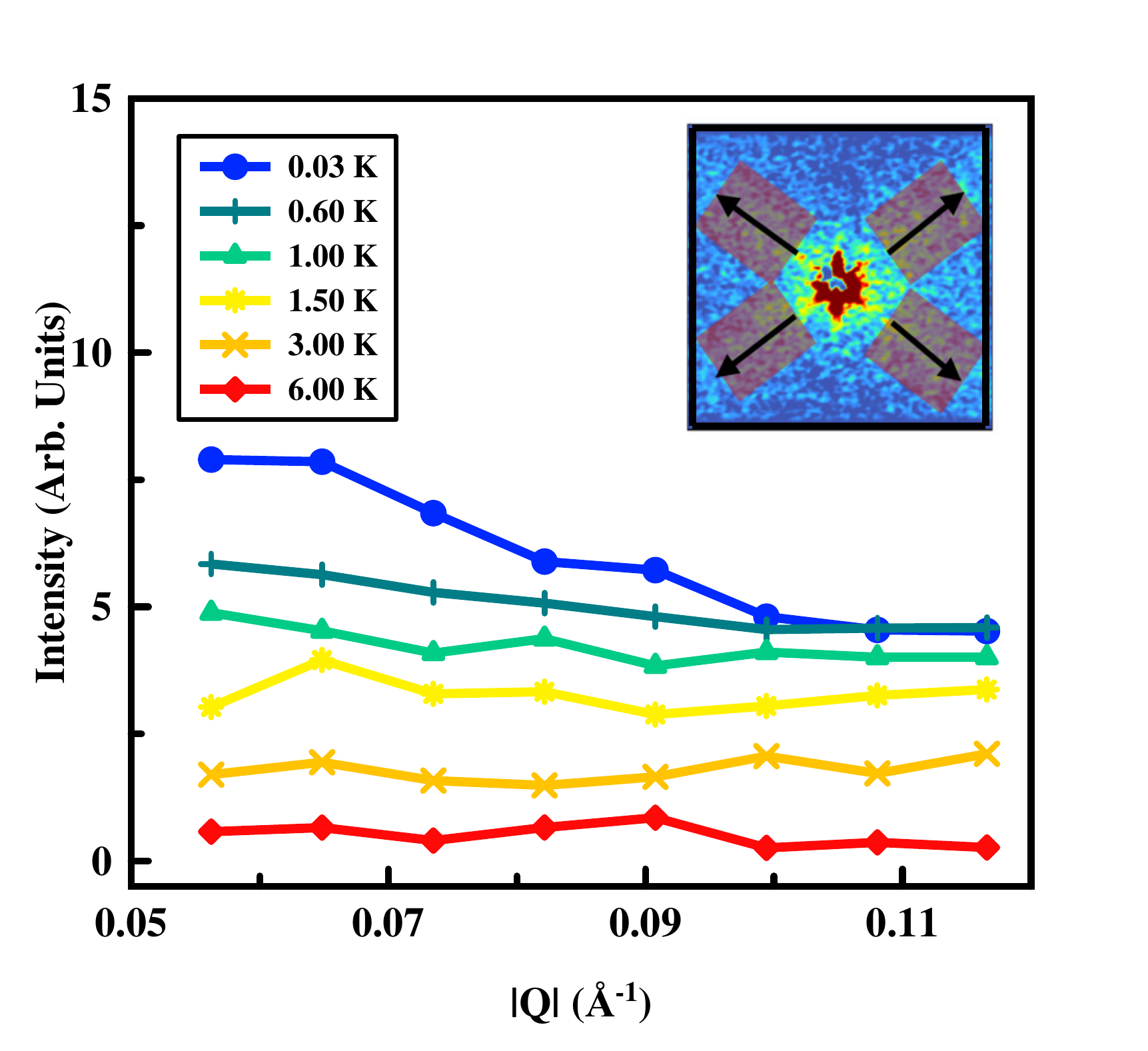}
\par
\caption{\emph{ $|Q|$ dependence of the $\langle 111 \rangle$  rods of scattering in \pyro{Yb}{Ti}.~ } Integrations of the scattering along these rods of scattering along the equivalent  [HHH] directions at different temperatures are shown. A high temperature, T=0.8 K data set, has been used as a background for all data shown.  The inset qualitatively shows the area of integration on the SANS detector. }
\label{fig:yb_qint}
\end{figure}

\begin{figure}[htb]
\linespread{1}
\par
\includegraphics[width=\linewidth]{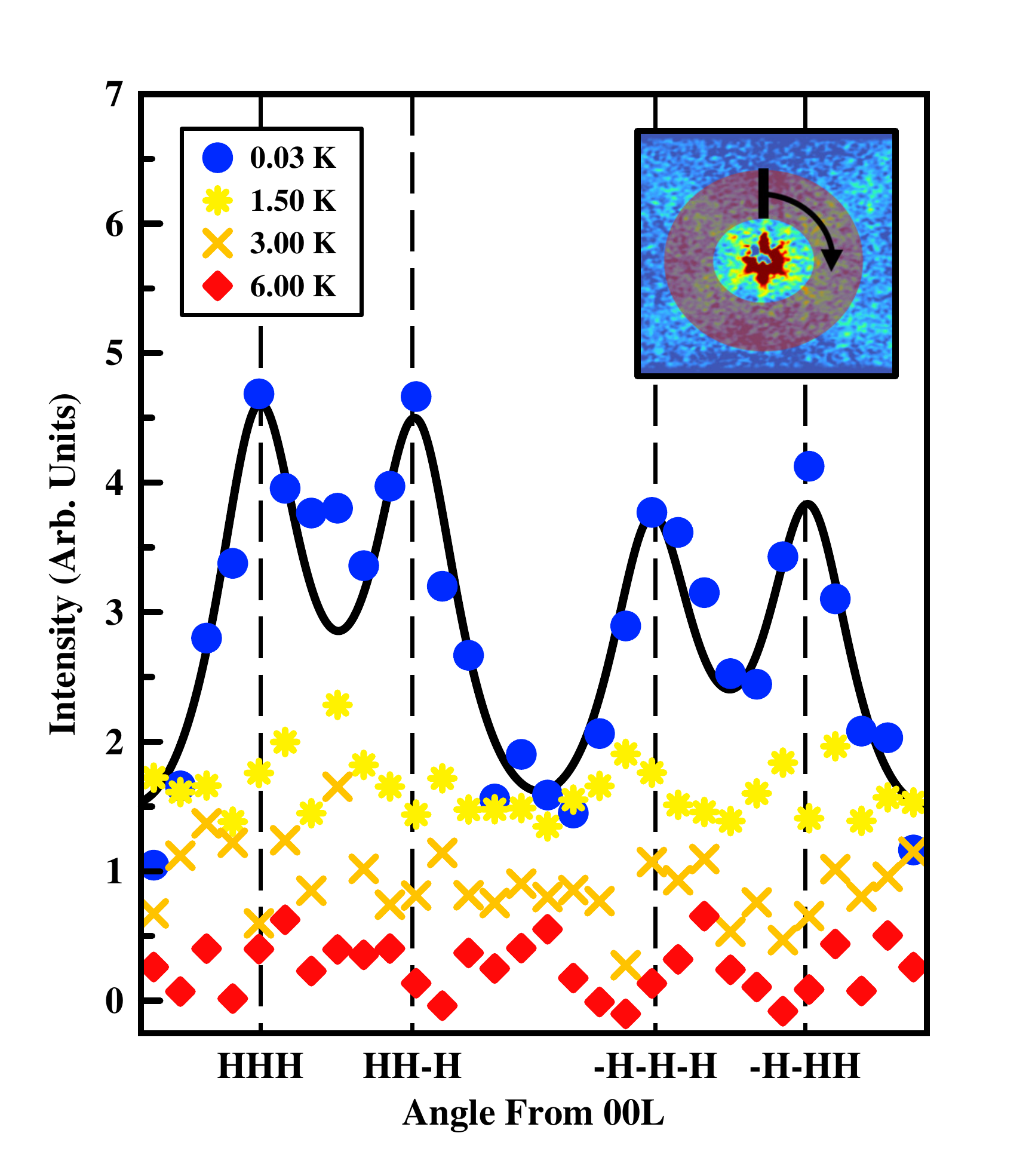}
\par
\caption{\emph{Angular or azimuthal dependence of the SANS data in \pyro{Yb}{Ti}.~ } The azimuthal integration of the SANS data over a fixed $|Q|$ range is shown. This azimuthal dependence peaks along the HHH directions below $\sim$ T=1 K. The T=0.03 K data has been fit to four Lorentzians, shown as the solid line, primarily to guide the eye. The HHH directions are shown as vertical dashed lines. The inset qualitatively shows the area of integration on the SANS detector, with the black bar serving as a reference for the zero of the angle. }
\label{fig:yb_qazi}
\end{figure}

The temperature dependence of the SANS from single crystal \pyro{Yb}{Ti} in zero magnetic field has been measured over a very large dynamic range, $\sim$ 0.03 to 8 K, allowing the diffuse scattering and quasi-Bragg scattering to be studied from temperatures much less than T$_C$ to much greater than $\Theta_{CW}$.  Below T=1 K, these data sets were measured on cooling only, starting from T=1 K.  For this reason it is the lower T$_C$ from the hysteresis at the phase transition that would be relevant.  The broad evolution of the SANS signal with temperature can be appreciated from Figs.~\ref{fig:yb_qint} and~\ref{fig:yb_qazi}.  Fig.~\ref{fig:yb_qint} shows the $|Q|$ dependence of the integrated rods of scattering along $\langle 111\rangle$ , as shown in the inset to Fig.~\ref{fig:yb_qint}.  Note that the magnetic form factor appropriate to Yb$^{3+}$ has no $|Q|$ dependence over the range of  $|Q|$ relevant to the SANS measurements.

Fig.~\ref{fig:yb_qint} shows that the SANS signal is flat in $|Q|$, but rising with decreasing temperature down to $\sim$ 1 K.  Below T=1 K, it shows clear $|Q|$ dependence increasing to low $|Q|$.  This is particularly pronounced below T$_C$, with the interpretation that this $|Q|$ dependence arises from the large $|Q|$ tail of the quasi-Bragg peaks.  Fig.~\ref{fig:yb_qazi} shows the angular or azimuthal dependence of the diffuse SANS signal over a fixed finite $|Q|$ range for selected temperatures, as indicated in the inset to Fig.~\ref{fig:yb_qazi}.  One can see that at high temperatures, above $\sim$ 1 K, there is no azimuthal dependence to the SANS signal although the overall strength of the {\bf Q}-independent SANS signal increases with decreasing temperature.  However at temperatures below T=0.50 K  $\sim$ $\Theta_{CW}$, a pronounced azimuthal dependence develops, with four peaks observed centered on the four equivalent $\langle 111 \rangle$ directions accessible in the measurement.  

\begin{figure}[htb]
\linespread{1}
\par
\includegraphics[width=\linewidth]{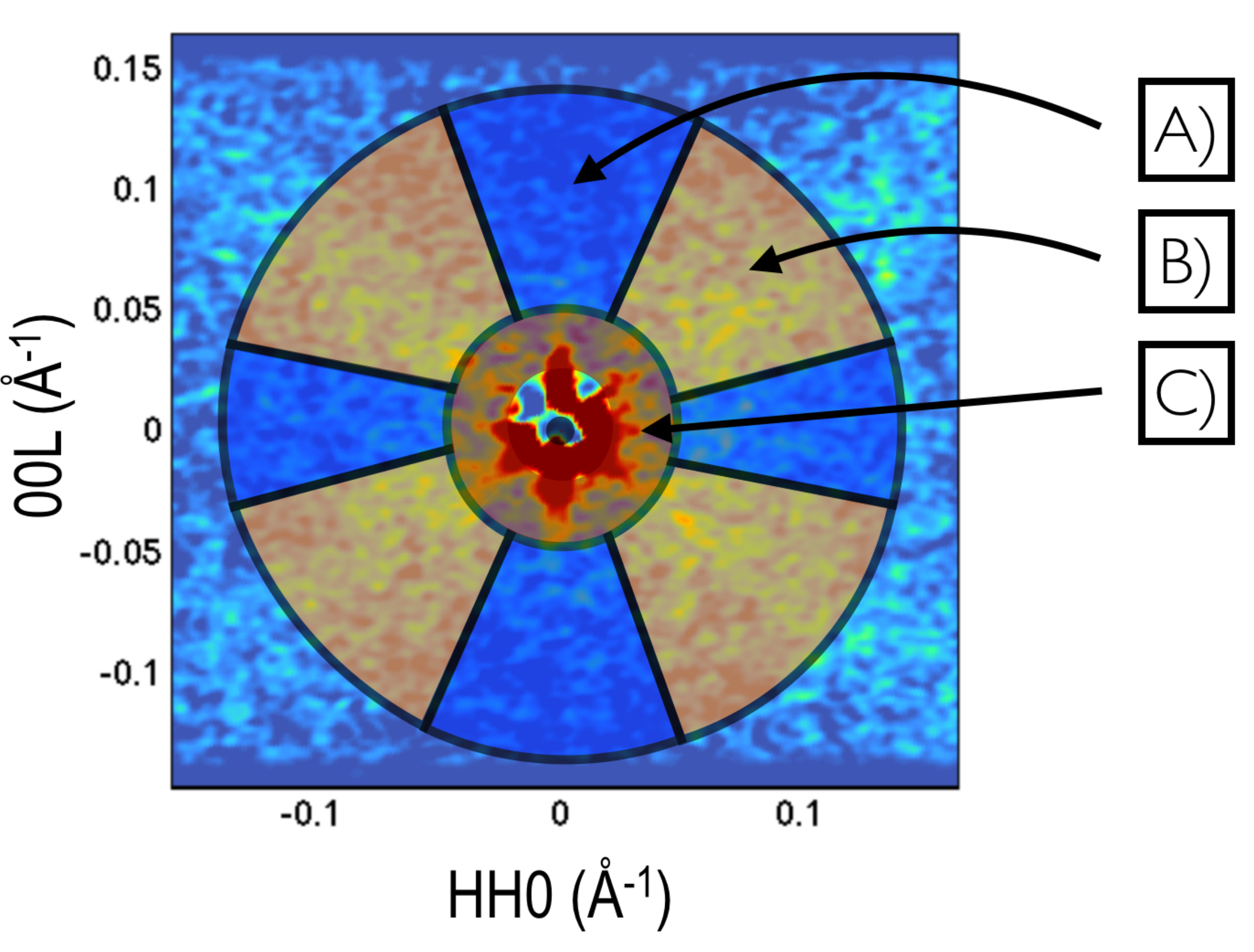}
\par
\caption{\emph{The areas of integration of the SANS data for \pyro{Yb}{Ti}.~ } A qualitative illustration of the regions of interest on the SANS detector are shown: Region A makes up the area of the (HHL) plane in between the $\langle 111 \rangle$ rods of diffuse scattering. Region B is centered on the $\langle 111 \rangle$ rods of diffuse scattering.  Region C corresponds to low $| Q |$ quasi-Bragg scattering. The region of the detector at very small $|Q|$, below $| Q |$=0.015 \AA$^{-1}$, is discarded due to potential contamination with the incident neutron beam.}
\label{fig:yb_select}
\end{figure}

To examine the details of the temperature dependence of the diffuse and quasi-Bragg scattering in \pyro{Yb}{Ti}, we divided the regions of our two dimensional SANS data sets as shown in Fig.~\ref{fig:yb_select}.  Region C covers an annular area near {\bf Q}=0, over which most of the quasi-Bragg scattering is distributed.  A small region near {\bf Q}=0 is excluded as it is potentially contaminated with the transmitted neutron beam (in principle, this should not matter as we employ a high temperature background subtraction; nonetheless it is safer to eliminate this very small {\bf Q} region from the analysis).  Region B covers pie-shaped regions extending at larger {\bf Q} along the four equivalent $\langle 111 \rangle$ directions, again as indicated in Fig.~\ref{fig:yb_select}.  This area covers the diffuse magnetic scattering on the $\langle 111 \rangle$ rods, distinct from the small {\bf Q} quasi-Bragg scattering.  Finally region A covers four pie-shaped regions, again at larger {\bf Q}, but now between the $\langle 111 \rangle$ rods of diffuse scattering.

\begin{figure}[htb]
\linespread{1}
\par
\includegraphics[width=\linewidth]{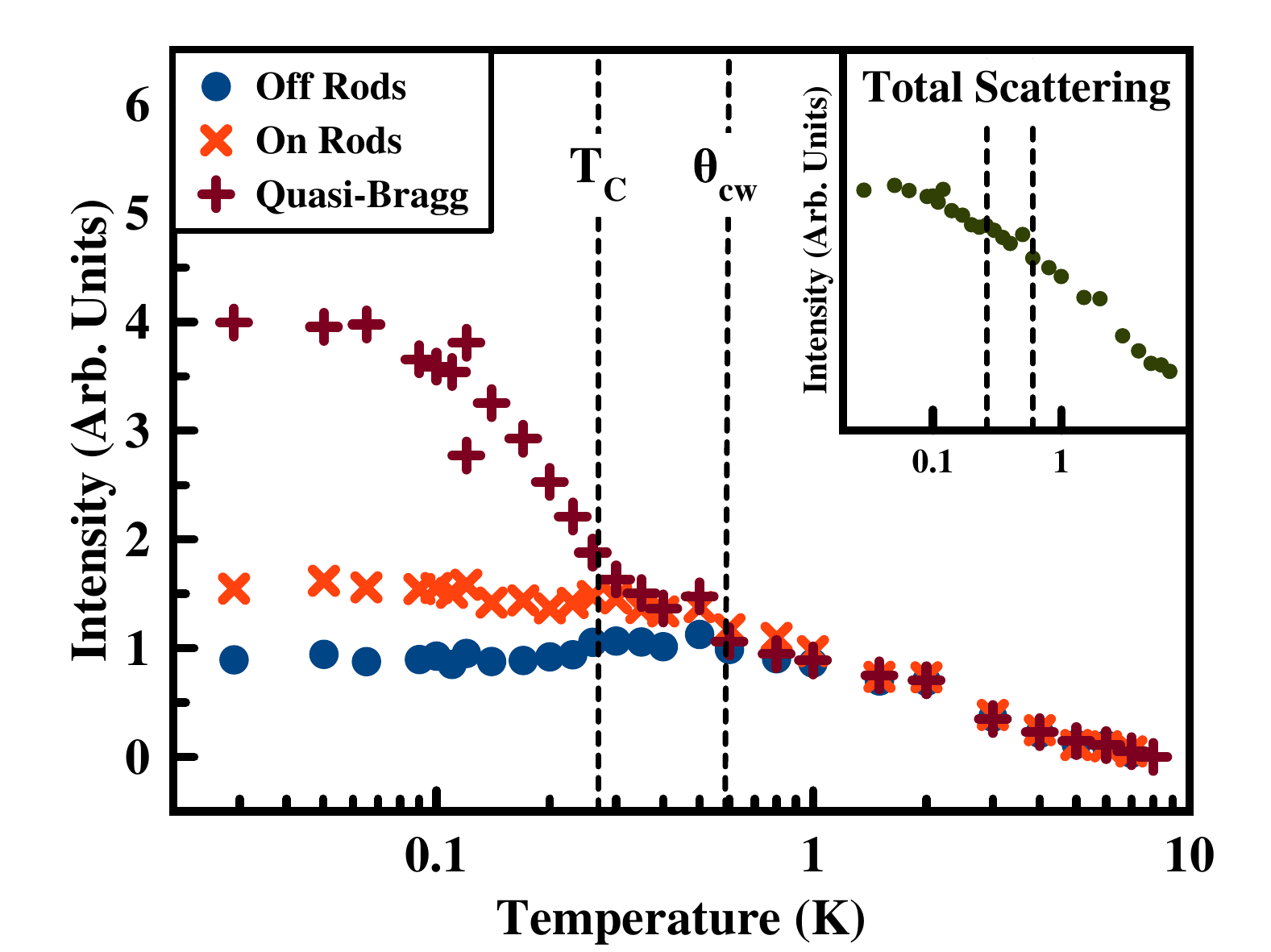}
\par
\caption{\emph{The temperature dependence of the three integrations of the SANS signal are shown along with fiducial markers indicating T$_C$=0.26 K and $\Theta_{CW}$=0.50 K for \pyro{Yb}{Ti}.~ } The integration of Fig.~\ref{fig:yb_select} region A indicates ``off-rod" scattering, that of Fig.~\ref{fig:yb_select} region B indicates ``on-rod" scattering and that of Fig.~\ref{fig:yb_select} region C indicates ``quasi-Bragg" scattering, all normalized to detector area, with a T =8 K data set subtracted. The inset shows the integration over all three sectors combined. Vertical lines in both the main and inset indicate T$_C$=0.26 K and $\Theta_{CW}$=0.50 K.}
\label{fig:yb_all}
\end{figure}

We can then look at the SANS signal normalized to detector pixel in each of these three areas - ``on-rod" diffuse scattering, ``off-rod" diffuse scattering, and quasi-Bragg peak scattering, as a function of temperature from just below T=0.03 K to just above T=8 K.  This is what is shown in Fig.~\ref{fig:yb_all}, with temperature on a logarithmic scale, as the dynamic range in temperature is $>$ 270.  One can see that above $\sim$ $\Theta_{CW}$, all three of the SANS intensities are essentially identical, meaning that the SANS signal is isotropic and flat, consistent with what is seen in the full data sets, eg. Fig.~\ref{fig:yb_panel} b).  However the ``off-rod" scattering peaks at T=0.50 K and decreases substantially below this temperature, with an inflection point near $\sim$ T=0.20 K.  The ``on-rod" scattering and quasi-Bragg scattering continue to increase below $\Theta_{CW}$, with the ``on-rod" scattering saturating below T=0.30 K.  The separation between the ``on-rod" scattering and the ``off-rod" scattering signifies the appearance of the rods of diffuse scattering along $\langle 111 \rangle$, and this happens clearly in the SANS data below T=0.50 K $\sim$ $\Theta_{CW}$.  The $\langle 111 \rangle$ rods of diffuse scattering are most apparent when the contrast between ``on-rod" scattering and ``off-rod" scattering is greatest, and this occurs below $\sim$ T=0.20 K, when both ``on-rod" and ``off-rod" scattering are saturated at their low temperature limits.

In contrast the quasi-Bragg scattering separates from the ``on-rod" diffuse scattering at $\sim$ 0.25 K $\sim$ T$_C$.  The growth of this quasi-Bragg scattering shows downwards curvature below T$_C$, consistent with typical order parameter behaviour.  This is in spite of the fact that this quasi-Bragg scattering is not due to true long range order, as we have already observed that magnetic field-induced ferromagnetic order shows no quasi-Bragg scattering, or any other SANS signal within our field of view, for that matter (see Fig.~\ref{fig:yb_panel} c).  The total SANS scattering on our detector, excluding a small region at very small {\bf Q} to avoid contamination with the transmitted beam (as discussed above) is shown as a function of temperature in the inset to Fig.~\ref{fig:yb_all}.  Both T$_C$ and $\Theta_{CW}$ are indicated on this plot, and it is clear that the overall SANS signal increases monotonically with temperature.  There appears to be a change in slope of the overall SANS signal near $\Theta_{CW}$, but there is no clear indication of T$_C$ $\sim$ 0.26 K.  The only clear indication of T$_C$ in the SANS data is the onset of the isolated, strong quasi-Bragg scattering shown in the main panel of Fig.~\ref{fig:yb_all}.

\begin{figure}[htb]
\linespread{1}
\par
\includegraphics[width=\linewidth]{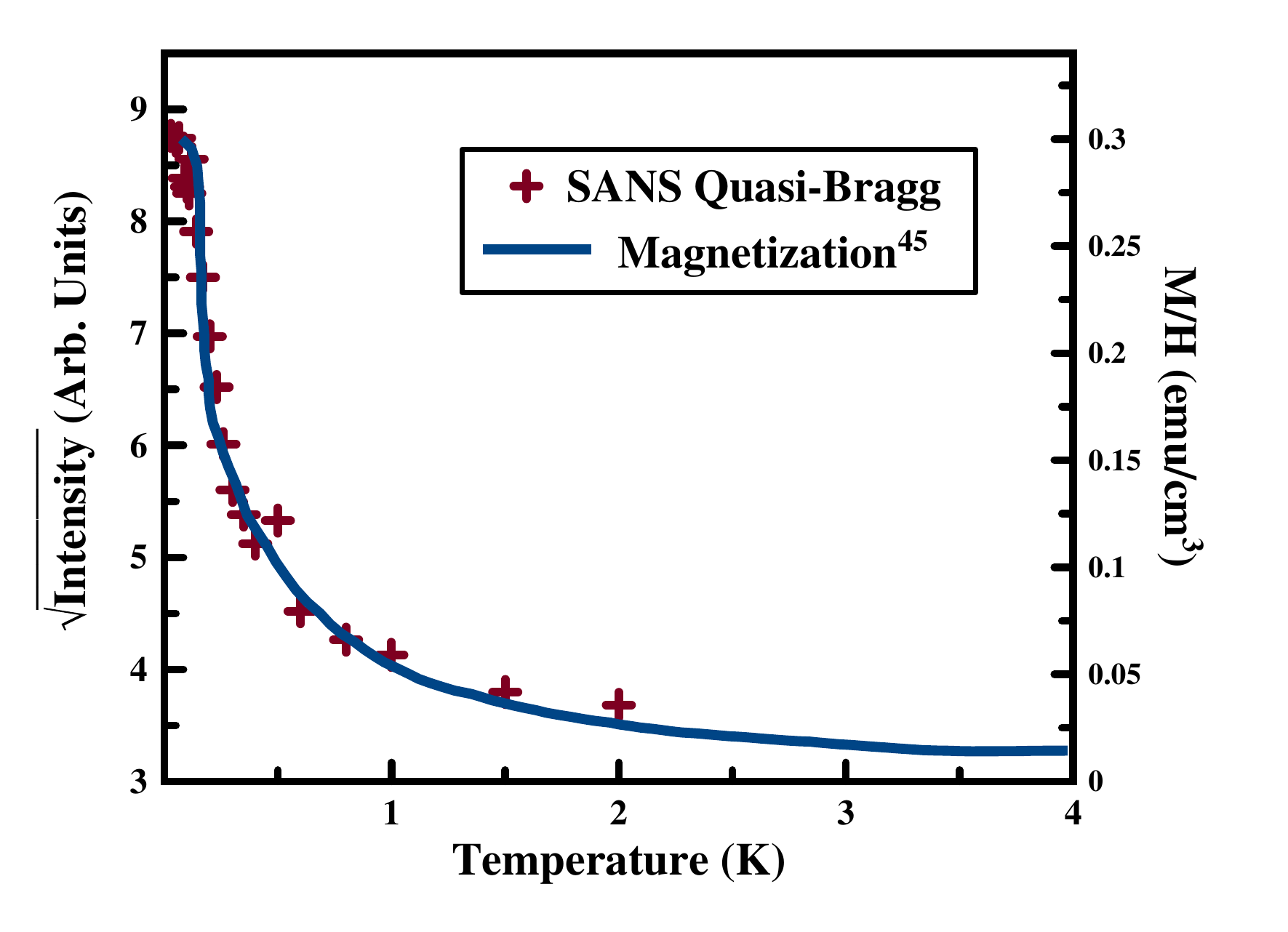}
\par
\caption{\emph{A comparison of the temperature dependence of the near ${\bf Q}$=0 quasi-Bragg scattering and the magnetization in \pyro{Yb}{Ti}.~ }  The square root of the SANS quasi-Bragg signal is plotted with reference to the left y-axis, while the solid line corresponds to single crystal magnetization data in 5 to 100 Oe field applied along the [100] direction with reference to the right y-axis, adapted from Lhotel et al.~\cite{lhotel2014}}  
\label{fig:yb_mag}
\end{figure}

It is interesting to compare our quasi-Bragg scattering to a direct measurement of the magnetization of single crystal  \pyro{Yb}{Ti}, as the magnetization is the order parameter for a ferromagnetic phase transition, while Bragg scattering from long range order should scale as the square of the order parameter.  In Fig.~\ref{fig:yb_mag} we show a comparison between the $\sqrt{}$ (quasi-Bragg intensity) vs temperature on a linear scale, plotted against the temperature dependence of the single crystal magnetization measured with a 5 Oe magnetic field applied along the [100] direction below 1~K and 100 Oe between 1~K and 4~K measured and reported by Lhotel et al.~\cite{lhotel2014}.  This magnetization data also shows weak hysteresis near T=0.15 K, with a separation of $\sim$ 0.01 K between warming and cooling curves (not evident in this plot).  As can be seen from Fig.~\ref{fig:yb_mag} the proportionality of quasi-Bragg scattering and magnetization is very close, despite the fact that the single crystal sample that was the subject of the magnetization study has an anomaly in C$_P$ just below 0.2 K, while the single crystal that was the subject of the SANS study shows a primary C$_P$ anomaly at 0.26 K and a secondary anomaly at 0.18 K, as shown in Fig.~\ref{fig:yb_heatcap}.  

The extent of the quasi-Bragg scattering in $|Q|$, $\sim$ 0.05 \AA$^{-1}$, as well as a simple Guinier analysis of the SANS signal, are consistent with the finite extent of the ferromagnetic domains giving rise to the quasi-Bragg scattering of $\sim$ 140~$\pm$ 30 \AA~ at the lowest temperatures measured, or about 14 cubic unit cells at T $\sim$ 0.04 K.  Earlier work on the magnetic structure of \pyro{Yb}{Ti} below T$_C$, in those crystals where it was observed, show that the magnetically ordered state is a splayed ferromagnet~\cite{Gaudet2016,Yaouanc2016,Scheie2017,kermarrec2017ground,pecanha2017magnetic}, with most studies reporting a structure with moments pointing almost parallel to $\langle 100\rangle$ directions~\cite{Gaudet2016,Scheie2017,kermarrec2017ground,pecanha2017magnetic}.  The fact that strong diffuse rods of scattering co-exist with the quasi-Bragg scattering even at the lowest temperatures, suggest that the diffuse rods of scattering arise from the boundary regions between the domains.  As this diffuse scattering is rod-like along $\langle 111 \rangle$ directions, the boundaries correspond to sheets of spins normal to $\langle 111 \rangle$ directions.  

The pyrochlore structure is known to be described as a stacking of two dimensional triangular and kagome planes along the $\langle 111 \rangle$ direction.  Therefore, the rods of scattering could arise from spins within triangular or kagome planes, or a bilayer of one triangular and one kagome plane, normal to $\langle 111 \rangle$ directions, which bound the $\sim$ $\langle 100\rangle$ ferromagnetic domains.  These domain boundaries must be much thinner along $\langle 111 \rangle$ directions compared to normal to this direction, in order to manifest themselves as rods of scattering.  In addition the moments within the domain boundaries must themselves lie within the kagome and triangular planes, in order to contribute to the scattering.  This is because the un-polarized neutron scattering cross section is sensitive only to components of moment lying in a plane perpendicular to {\bf Q}.  For consistency with earlier measurements of the correlation length normal to the rods of diffuse scattering~\cite{ross2011dimensional}, the magnetic moments within the domain boundaries would correspond to ferromagnetic patches of spins oriented within the kagome and triangular planes.  Such a short ranged, ferromagnetic domain wall spin texture would be intermediate between the spin arrangement of two neighbouring $\langle 100\rangle$ ferromagnetic domains, and the domain boundaries would then play the role of alleviating the exchange energy penalty associated with an abrupt transition between two ferromagnetic domains with approximately orthogonal local $\langle 100\rangle$ polarizations. 
  
\subsection{\pyro{Ho}{Ti}}

\begin{figure*}[htbp]
\linespread{1}
\par
\includegraphics[width=\linewidth]{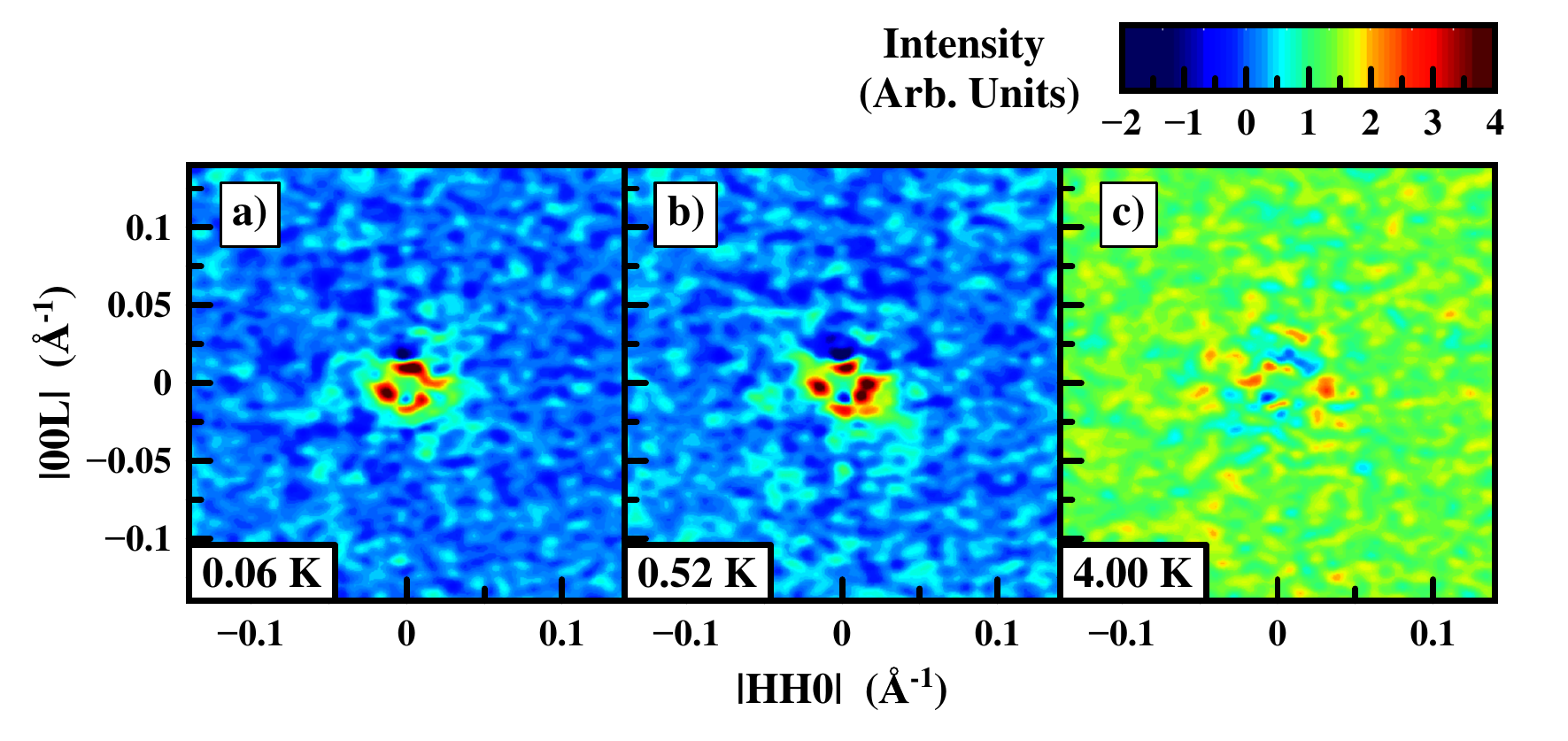}
\par
\caption{\emph{ Small angle neutron scattering data of \pyro{Ho}{Ti} at three characteristic temperatures.~}  a)  SANS data is shown at T=0.06~K, well below the spin freezing temperature, b) Data is shown at T=0.52~K just below the spin freezing temperature and well below the spin ice transition. c) Data is shown at T=4~K, above the spin ice transition. All data sets have had a high temperature, T=10~K, data set subtracted from them. }
\label{fig:ho_panel}
\end{figure*}

\begin{figure*}[htbp]
\linespread{1}
\par
\includegraphics[width=\linewidth]{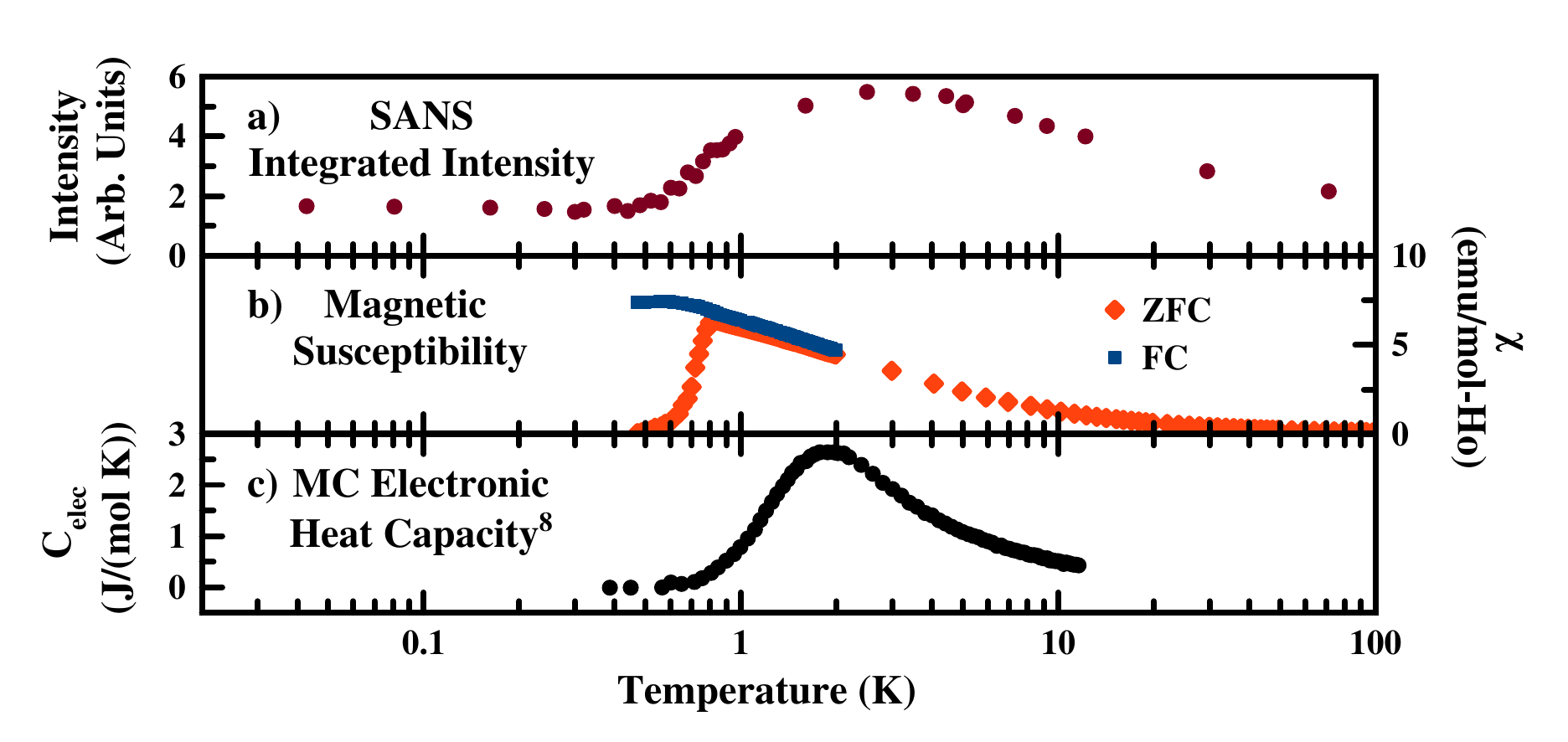}
\par
\caption{\emph{A comparison of SANS data, magnetic susceptibility, and electronic heat capacity in \pyro{Ho}{Ti}.~ } a) Integrated intensity of SANS data with near beam stop scattering removed. b) ZFC/FC susceptibility of \pyro{Ho}{Ti} powder, field of 50 Oe c) Montecarlo simulated heat capacity of \pyro{Ho}{Ti} with nuclear contribution subtracted adapted from Bramwell et al.~\cite{Bramwell2001a} Note that the peaked behaviour in a) is matched well by behaviour in c) at T=2 K.}
\label{fig:ho_all}
\end{figure*}

Representative SANS data sets from single crystal \pyro{Ho}{Ti} are shown in Fig.~\ref{fig:ho_panel} for a) T=0.06 K, b) T=0.52 K and c) T= 4 K, all in zero magnetic field and with a T=10 K background SANS data set subtracted from them.  The magnetic contribution to C$_P$ appropriate for \pyro{Ho}{Ti} is hard to determine experimentally, due to the presence of a large nuclear Schottky anomaly at low temperatures~\cite{Bramwell2001a}.  Nonetheless, the magnetic C$_P$ for \pyro{Ho}{Ti} as a function of temperature has been estimated on the basis of Monte Carlo modelling of a suite of experimental data, and this is adapted from Bramwell et al.~\cite{Bramwell2001a} here in Fig.~\ref{fig:ho_all} c).   The magnetic susceptibility of powder samples of \pyro{Ho}{Ti} measured in a 50 Oe magnetic field is shown in Fig.~\ref{fig:ho_all} b).  SANS data in Fig.~\ref{fig:ho_panel} a) and b) taken at both T=0.06 K and T=0.52 K should therefore be well within the spin ice regime for \pyro{Ho}{Ti}, where very little magnetic C$_P$ remains.  The magnetic entropy of \pyro{Ho}{Ti} at these temperatures is $\sim$ that of the Pauling residual entropy for spin ice~\cite{ramirez1999zero,pomaranski2013absence}.  The data set at T=0.52 K is also just below the break between the field-cooled and zero-field cooled susceptibilities that signifies spin freezing and the out-of-equilibrium nature of this spin ice state below $\sim$ 0.6 K in \pyro{Ho}{Ti}.  In contrast, T= 4 K is above the broad maximum in the magnetic C$_P$ for \pyro{Ho}{Ti}, and at these temperatures \pyro{Ho}{Ti}  is in more of a conventional paramagnetic state, with a proliferation of classical magnetic monopoles~\cite{fennel2009coulomb}.

The SANS signal in single crystal \pyro{Ho}{Ti} shown in all panels of Fig.~\ref{fig:ho_panel} displays no azimuthal dependence at any temperatures.  It is clear nonetheless, that there is a significant decrease in the SANS scattering over the range of {\bf Q} covered in the measurement, with a much reduced SANS signal at low temperatures, compared with T=4 K.  This is qualitatively consistent with the expectation of the dipolar spin ice model.  The calculated S({\bf Q}) within the dipolar spin ice model differs from that calculated within the near neighbour spin ice model, primarily by a suppression of the scattering at small {\bf Q}~\cite{den2000dipolar,melko2001,Bramwell2001}, and this suppression will occur at low temperatures, below the broad C$_P$ maximum.

As there are no features in the SANS signal for single crystal \pyro{Ho}{Ti} beyond a uniform increase or decrease as a function of temperature, we can integrate over the full detector, excluding only a small region about {\bf Q}=0 to eliminate any extraneous contribution from the transmitted beam. This is shown in Fig.~\ref{fig:ho_all} a).  No background subtraction has been employed for this integration.  We once again characterize this SANS signal over a wide dynamic range in temperature from $\sim$ T=0.04 K to 220 K, a dynamic range of more than 500.  The temperature dependence of the SANS signal bears similarities to the magnetic C$_P$, in that they both peak at roughly the same temperature, T $\sim$ 3 K for the SANS signal and $\sim$ 2 K for the magnetic C$_P$.  The SANS signal is also broader in temperature than the magnetic C$_P$, especially on the high temperature side, where its temperature dependence better resembles that of the high temperature magnetic susceptibility.  The fall off of the SANS signal at low temperatures is roughly consistent between all three of the zero field cooled magnetic susceptibility, the magnetic C$_P$,  and the SANS signal.  All three are small or zero below $\sim$ 0.6 K. 

We therefore associate the low temperature SANS signal with monopole excitations corresponding to defects within the spin ice ground states, that is, 3 in 1 out and 1 in 3 out spin configurations on a tetrahedron.  These possess a relatively large net moment per tetrahedron, and clearly there are very few remaining in the system below $\sim$ 0.6 K.  The population of the monopole excitations peak between 2 and 4 K.  Beyond 4 K, additional high energy local defect structures (all in, all out configurations on a tetrahedron) mix in, diminishing the SANS signal, as these possess no net moment per tetrahedron.

\section{Conclusions}

We have presented single crystal SANS studies over a large dynamic range of temperatures of two members of the \pyro{RE}{Ti} family of cubic pyrochlore magnets that display net ferromagnetic Curie-Weiss susceptibilities.  \pyro{Yb}{Ti} is a candidate for a quantum spin ice state material.  While its SANS signal is isotropic at temperatures above $\sim$ 0.8 K, it displays three distinct  SANS signals within the (HHL) plane at lower temperature.  Diffuse rods of scattering along $\langle 111 \rangle$ directions develop at temperatures $\sim$ $\Theta_{CW}$ $\sim$ 0.5 K, as this ``on-rod" scattered SANS intensity separates from the ``off-rod" SANS intensity, which peaks near $\Theta_{CW}$.  Both the ``on-rod" and ``off-rod" SANS intensities plateau below $\sim$ T=0.2 K, and thus the $\langle 111 \rangle$ rods of diffuse scattering persist to the lowest temperatures measured. 

In addition, quasi-Bragg scattering is observed at very small {\bf Q}, with an onset close to T$_C$ $\sim$ 0.26 K, co-incident with the largest anomaly in the C$_P$ measured for this large single crystal.  This signal shows downwards curvature as a function of decreasing temperature, consistent with order parameter behaviour, and its square root correlates well with measurements of the low temperature magnetization previously measured on a different \pyro{Yb}{Ti} single crystal.  This quasi-Bragg SANS scattering is not evident at low temperatures in the presence of a strong [110] magnetic field, when the material is expected to exhibit true ferromagnetic long range order.  For this reason, the quasi-Bragg scattering is associated with finite size ferromagnetic domains polarized $\sim$ along $\langle 001 \rangle$ directions, and $\sim$ 140 \AA~ in extent.  The $\langle 111 \rangle$ rods of diffuse scattering are then most naturally associated with quasi-two-dimensional domain boundaries between these $\langle 100\rangle$ ferromagnetic domains, with the Yb moments polarized within the plane of the domain boundaries - so within the planes normal to $\langle 111 \rangle$ directions.  The $\langle 111 \rangle$ diffuse rods of scattering have been observed previously by several groups over a long period of time~\cite{bonville2004transitions,Ross2009,thompson2011}, as the ground state of \pyro{Yb}{Ti} in zero magnetic field has been long debated.  Nonetheless the physical origin of this diffuse scattering has been unclear.  We believe these observations clarify their nature. 

In contrast, SANS studies on single crystal \pyro{Ho}{Ti}, which is one of the prototypical magnets displaying a dipolar spin ice ground state, show an isotropic SANS signal at all temperatures measured.  Nonetheless the flat SANS intensity over the (HHL) plane shows pronounced temperature dependance, resembling the estimate for the magnetic C$_P$ in \pyro{Ho}{Ti} at least in terms of its low temperature behaviour.  At high temperatures, above the peak in C$_P$ or the SANS signal at $\sim$ 3 K, the SANS signal falls off more slowly with increasing temperature than the C$_P$, and better resembles the temperature dependence of the high temperature magnetic susceptibility.

\begin{acknowledgments}
This work was supported by NSERC of Canada and used resources at the High Flux Isotope Reactor , a DOE Office of Science User Facility operated by the Oak Ridge National Laboratory.  We wish to thank K. A. Ross, A. M. Hallas, S. Kuhn and M. J. P. Gingras for useful discussions. 
\end{acknowledgments}

\end{document}